    \documentstyle[proceedings,psfig]{crckapb}

\begin{opening}
\title{Abundances in Damped Lyman-alpha Systems and Chemical Evolution
of High Redshift Galaxies}

\author{Limin Lu}
\author{Wallace L.W. Sargent}
\institute{Caltech, Astronomy Dept., 105-24, Pasadena, CA 91125, USA}
\author{Thomas A. Barlow}
\institute{Caltech, IPAC, 100-22, Pasadena, CA 91125, USA}

\end{opening}

\runningtitle{Damped Lyman-alpha Abundances}

\begin{document}

\begin{abstract}
Recent abundance measurements in damped Ly$\alpha$ 
galaxies, supplemented with unpublished Keck observations, are discussed.
The metallicity distribution with cosmic time is examined for clues
about the degree of enrichment, the onset of initial star formation,
and the nature of the galaxies.
The relative abundances of the elements are compared with the
abundance patterns in Galactic halo stars and in the Sun, taking
into account the effects of dust depletion, in order to 
gain insight into the stellar processes and the time scales
by which the enrichment occurred.   
\end{abstract}

\section{Introduction}
 
Without doubt, one of the most significant developments over the last
decade in quasar absorption line studies has been the recognition
of high-redshift damped Ly$\alpha$ (DLA) absorption systems 
as the progenitor of present-day galaxies (Wolfe et al 1986) and
of their utility for {\it direct} observational study of the
chemical enrichment history of normal galaxies since the very early
epoch. Such studies are therefore complementary to
the traditional approach to galactic chemical evolution based on
studies of stellar populations in the Milky Way and in 
nearby galaxies (these proceedings). Lauroesch et al (1996) summarized
then-existing DLA abundance measurements and discussed their
implications for understanding the chemical evolution history of
normal galaxies. Results from two new surveys have since appeared
in press (Lu et al 1996, hereafter LSBCV; Pettini et al 1997a,b), which represent
a significant improvement in quantity and quality over previous 
measurements.  The Pettini et al studies contain the most 
measurements of Zn and Cr abundances obtained from high S/N medium
resolution (FWHM=30-80 km/s) observations. The Lu et al study was based on 
high S/N echelle (FWHM=6-8 km/s) observations collected with the Keck~I
10m telescope and comprises the single largest source of abundance 
measurements for other elements (C, N, O, Si, S, Mn, Fe, Ni,  as well
as Zn and Cr). Our discussion will focus on the new survey results
and how they provide new insight on understanding early galactic chemical
evolution.

Excellent discussions of various uncertainties that may affect 
the DLA abundance measurements and interpretations can be found
in Lauroesch et al (1996). 
Since the LSBCV and Pettini et al (1997a,b) surveys 
used weak lines to derive column densities and
elemental abundances, uncertainties resulting from saturated absorption
lines should be minimal. Typical measurement uncertainties in the 
abundances are $\sim 0.1$ dex.  More precise determinations of
oscillator strength ($f$-value) for several important species
(e.g., Si~\textsc{ii}, Fe~\textsc{ii}, Cr~\textsc{ii}) have become
available recently (see Table 2 of Savage \& Sembach 1996 and 
references therein); the new surveys have adopted these more
accurate $f$-values.  The abundance estimates generally assume
that the bulk of the absorbing H gas is neutral and that column 
density ratios of dominant ions in neutral gas to \textsc{H~i}
yield direct measures of the actual gas-phase abundance of the
elements with 
negligible ionization corrections. This assumption appears to be
corroborated by simple ionization models (Viegas 1995, Lu et al 1995;
Prochaska and Wolfe 1996).\footnote{Green et al (1995) reported a large
ionization correction for the $z=1.77$ DLA system toward Q 1331+170.
Since the claim was based on a N(\textsc{O~i})/N(\textsc{S~ii})
ratio from saturated \textsc{O~i} absorption, the inferred large
ionization correction is somewhat suspect.}

\section{Distribution of Metallicity as a Function of Redshift}


\begin{figure}
\centerline{\vbox{
\psfig{figure=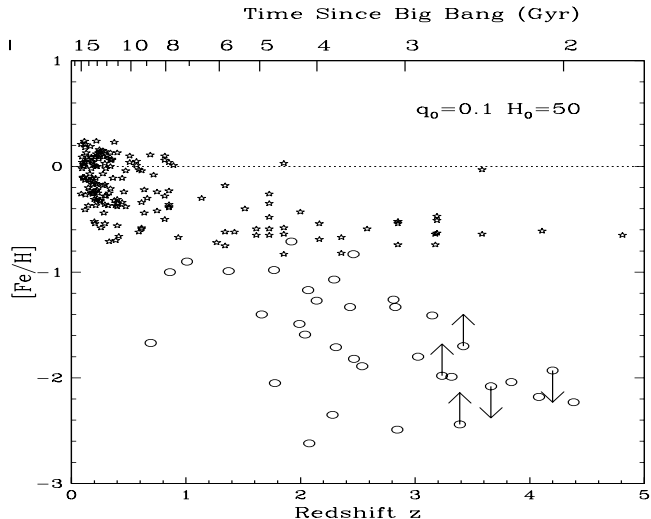,height=3.0in,width=3.5in}
}}
\caption[]{[Fe/H] vs $z$ for damped Ly$\alpha$ systems (open circles) and for
Galactic disk stars (star symbols; Edvardsson et al 1993), 
where [Fe/H]$\equiv$log(Fe/H)$-$log(Fe/H)$_{\odot}$.
The stellar ages have been converted into redshifts for a $q_0=0.1$ and $H_0=50$
cosmology.}
\end{figure}

Figure 1 shows the distribution of [Fe/H] vs $z$ for the sample of
DLA galaxies (open circles).
The ``$\star$'' symbols map out  the similar relation for a sample of
Galactic disk stars in the solar neighborhood (Edvardsson et al 1993).
The DLA data are taken from Table 16 of LSBCV (see LSBCV for references)
with the addition of
14 new measurements based on unpublished work of our group and
of Wolfe and Prochaska (5 systems).
The addition of 7 new measurements at $z>3$ considerably improves the
statistics in this redshift range.  Some of the new measurements
are still preliminary and may be subject to change
in the final analysis. However, the changes (if any) are expected to
be small and should have no effect on any of the conclusions.
Most existing measurements are for $z>1.6$; extending the programs
 to lower redshifts clearly remains a priority. 
Below we summarize the main conclusions.

1. Typical DLA galaxies have $-2.5<$[Fe/H]$<-1$,
corresponding to 1/300 to 1/10 solar metallicity.
The N(\textsc{H~i})-weighted mean metallicity is $<$[Fe/H]$>\simeq-1.5$
at $\langle z\rangle=2.5$. The
low metallicities are consistent with them being young galaxies in the
early stages of chemical enrichment. If DLA systems eventually
evolve to solar mean metallicity at $z=0$, the low metallicities
at $z>2$ suggest that most of the baryonic matter in the galaxies 
should be in the gas phase rather than in stars and that most of the
star formation should occur at $z<2$, consistent with results from
deep galaxy redshift surveys (cf, Connolly et al 1997).
The N(H~\textsc{i})-weighted mean metallicity
in terms of Zn is about a factor of 2-3 (0.4 dex) higher (Pettini et al 1997b),
possibility suggesting that some of the Fe atoms are locked up in
dust grains in the DLA galaxies (section 3).

2.  DLA galaxies appear significantly
less metal-enriched than the Galactic disk in its past. Taken at face value,
this argues against the suggestion that DLA galaxies
are  high-redshift (proto-) galactic disks (Wolfe 1988).
Rather, the metallicities of the DLA systems
are more similar to those of globular clusters and nearby
dwarf galaxies. This result may be in potential conflict with
the evidence that DLA systems appear to show kinematics
characteristic of fast-rotating ($v_{rot}\sim 250$ km/s) disks 
(Prochaska \& Wolfe 1997).
However, the rotating disk interpretation of
the DLA kinematics (Prochaska \& Wolfe 1997) may not be the only possible
explanation (Haehnelt, Steinmetz, \& Rauch 1997).
The significance of the metallicity discrepancy between the
DLA galaxies and the Milky Way disk may also be questioned for several reasons.
(1) Age determinations for old stars can be quite uncertain. The ages of
the stars in Figure 1 at $z>2$ carry an uncertainty of 2-3 Gyrs from
measurement errors alone, and there may also be systematic effects 
(Edvardsson et al 1993).
(2) The true metallicities of DLAs may be
higher than that indicated by [Fe/H] if significant dust depletion
of Fe has occurred (cf, Pettini et al 1997a). 
However, this effect alone cannot
explain the discrepancy entirely because the inferred depletion of Fe
is only $\sim 0.4$ dex (section 3).
(3) DLA systems may preferentially probe regions of disk galaxies
(assuming they exist at the relevant redshifts) beyond the
equivalent of the solar circle owing to the larger
absorption cross section.
The metallicity gradient known to exist in local
spiral disks (Vila-Costas \& Edmunds 1992) 
then helps to explain the discrepancy (Ferrini, Molla, \& Diaz 1997).
 
3. The mean metallicity of DLA systems clearly increases 
from $z>4$ to $z\sim 2$ as expected. 
However, there is a factor of $\sim$30 scatter
in [Fe/H] at $2<z<3$, presumably reflecting differences in the
formation epoch/star formation history of the galaxies and/or a mixture
of morphological types.  LSBCV found  that the mean metallicity of DLAs
increases fairly abruptly at $z<3$ compared to $z>3$, which could
be interpreted to signal the onset of star formation in DLA galaxies.
The much more uniform sampling of the $z>3$ region now
available shows that the change in metallicity from
$z<3$ to $z>3$ is actually much smoother. The metallicity distribution
appears to reach  a ``plateau'' value of [Fe/H]$\sim -2$ to $-2.5$ at $z>4$.
Coincidentally, this ``plateau'' metallicity
is identical (within the measurement uncertainties) to that  found for the
intergalactic medium (IGM) clouds at similar redshifts, as inferred
from the \textsc{C~iv} absorption
associated with Ly$\alpha$ forest clouds (Cowie et al 1995; Tytler et al 1995;
Songaila \& Cowie 1996).
This coincidence
suggests that the metals in DLA galaxies with [Fe/H]$\sim -2$ to $-2.5$
may simply reflect those in the IGM, possibly produced by Pop III stars
(Ostriker \& Snedin 1996).
If this interpretation is correct, then significant star formation did
not start in DLA galaxies until $z\sim 3-4$.
Such an inference is consistent with the decline in the neutral gas
content of DLA systems at $z>3$ (Storrie-Lombardi, McMahon, \& Irwin 1996), 
presumably because DLA galaxies
are still being formed at such high redshifts; and with the rapid decline
in the space density of quasars at $z>3$ (Schmidt, Schneider, \& Gunn 1995).
It will be important to study more DLA systems at the highest redshift possible
to confirm the reality of the ``plateau'' metallicity, and to improve
the accuracy of the metallicity determination for IGM clouds, which
at present may be uncertain by as much as a factor of 10.

\begin{figure}
\centerline{\vbox{
\psfig{figure=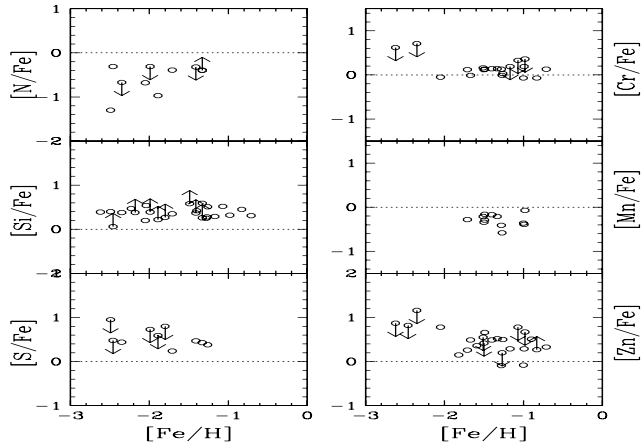,height=2.6in,width=3.5in}
}}
\caption{Relative abundance patterns of damped Ly$\alpha$ systems.}
\end{figure}

\section{Relative Abundances}

Abundance ratios of elements contain important clues to the
stellar processes (SN II, SN Ia, AGB stars)
and the time scales by which the elements were produced
(Wheeler, Sneden, \& Truran 1989). The straightforward 
conversion from redshift to cosmic time 
provides an accurate clock to gauge the progress
of SN II, SN Ia, and AGB star enrichment, which in turn may be
used to calibrate the events that affected Galactic chemical evolution.
However, unlike stellar abundances, the abundances derived from
interstellar absorption lines may not reflect the total 
abundance\footnote{We use {\it total abundance} or {\it intrinsic
abundance} interchangeably to indicate the abundance
of an element in both the gas phase and dust grains.}
of the elements since most elements are prone to condensation
onto dust grains (cf. Jenkins 1987). Consequently, possible dust depletion
effects must be taken into consideration in order to reach
reliable conclusions.

The abundance ratios to Fe of selected elements are plotted against
[Fe/H] of the systems in Figure 2, which represents an updated
version of Figure 23 in LSBCV incorporating our unpublished
measurements. Other elements are not shown either because
of too few measurements (e.g., C, O, Al) or suspicions of inaccurate
$f$-values (Ni; see end of this section).  
No significant evidence is found for the
abundance ratios to change with redshift over $z=1.6-4.4$.
The DLA abundance patterns are clearly different from the solar pattern.
Since the DLA galaxies at $z=2-4$ are at an epoch when the Galactic
halo was formed, they may be expected to show similar intrinsic abundance
patterns to those observed in halo stars. Indeed,
the DLA abundance patterns are very similar to 
that seen in Galactic halo stars\footnote{In this
discussion, Galactic {\it halo stars} refer to those with [Fe/H]$<-1$.}
and globular clusters (Wheeler et al 1989), showing a low N/Fe ratio, an overabundance
of Si and S relative to Fe (i.e., enhancement of $\alpha$-process
elements over Fe-group elements), and an underabundance of Mn relative
to Fe (i.e., the odd-even effect). The similarities suggest that
metal productions in most of the DLA systems under study were probably
dominated by massive stars via SN IIs. However, 
the super-solar Zn/Fe ratios in DLA systems,
$<$[Zn/Fe]$>\simeq 0.4$ dex,
are inconsistent with the abundance patterns seen in halo stars,
where Zn/Fe$\simeq$solar for all stars with [Fe/H]$>-3$ 
(Wheeler et al 1989).
Super-solar Zn/Fe ratios are often found in Galactic ISM clouds because,
while Zn is relatively unaffected by dust,
80-99\% of the Fe atoms are usually removed from the gas
phase by condensation onto dust grains (Jenkins 1987). 
The super-solar Zn/Fe ratios in DLAs
are likely caused by the same dust depletion effect
(Meyer \& Roth 1990; Pettini et al 1994;1997a).


Many attempts have been made to compare the DLA abundance patterns with
the solar pattern and the halo-star pattern for consistency, assuming
dust grains in DLA systems are the same as Galactic dust
(LSBCV; Lauroesch et al 1996; Kulkarni, Fall, \& Truran 1997;
Welty et al 1997; Vladilo 1997). Despite the efforts, 
however, no clear picture has emerged. LSBCV noted the close resemblance
between the DLA abundance patterns and the halo-star pattern but found
it difficult to interpret the ``anomalous'' Zn/Fe ratios in DLA systems 
as the consequence of dust depletion since it would have problem
explaining some other abundance ratios (e.g., N/Fe, Mn/Fe); they suggested that
the supersolar Zn/Fe ratios may be intrinsic to the nucleosynthesis.
Lauroesch et al (1996) also found evidence for a halo-star-like
abundance pattern and argued that, since the
magnitudes of nucleosynthesis effects and dust depletion effects can be
comparable, an unambiguous statement concerning the extent of
dust depletion in DLA systems is not yet possible. 
Welty et al (1997) found similar evidence for a halo-star-like pattern
and concluded that the DLA abundance patterns are affected by both 
nucleosynthetic effects and dust 
depletion - to different degrees for different systems. 
Kulkarni et al (1997), on the other hand, concluded that the DLA
abundances are equally well explained with a solar+dust depletion or a halo-star
pattern, but acknowledged that neither provides a perfect match to
the data.  Vladilo (1997) took a somewhat different approach.
Assuming DLAs have solar intrinsic Zn/Fe and Zn/Cr ratios and
contain Galactic-type dust, he used the
observed Zn/Fe or Zn/Cr ratios to infer the dust-to-metal ratio
appropriate for each DLA system which then allows him to predict the
amount of each element this is in dust grains.
The inferred total abundances of the $\alpha$-elements and Fe-group
elements showed abundance ratios within 0.2 dex of the solar ratio
(see Figure 1 of Vladilo 1997),
thus suggesting no evidence for $\alpha$-element enhancement
over Fe-group elements that is characteristic of halo stars. 

One common finding by all studies was that DLAs 
do not exhibit the heavy dust depletion seen in Galactic cool 
disk clouds. When dust
was advocated, the inferred depletion pattern was more similar to that
seen in warm disk clouds or warm halo clouds, where the degree of dust
depletion is much less than that observed in cool disk clouds
(Savage \& Sembach 1996).
 Welty et al (1997) also found that DLA abundance
patterns are similar to that seen in the ISM of the SMC.
The dust-to-metal ratios inferred for the DLA systems
are typically 0.4-0.8 times the Galactic value if the depletion
pattern in warm disk clouds is adopted as the reference, and the
corresponding dust-to-gas ratio is in the range of 2-25\% of
the Galactic value (Kulkarni et al 1997; Pettini et al 1997a; Vladilo 1997).
These values are similar to that found by Pei, Fall, \& Bechtold (1991)
based on a comparison of the colors of quasars with and without
DLA absorption in their spectra.


\begin{figure}
\centerline{\vbox{
\psfig{figure=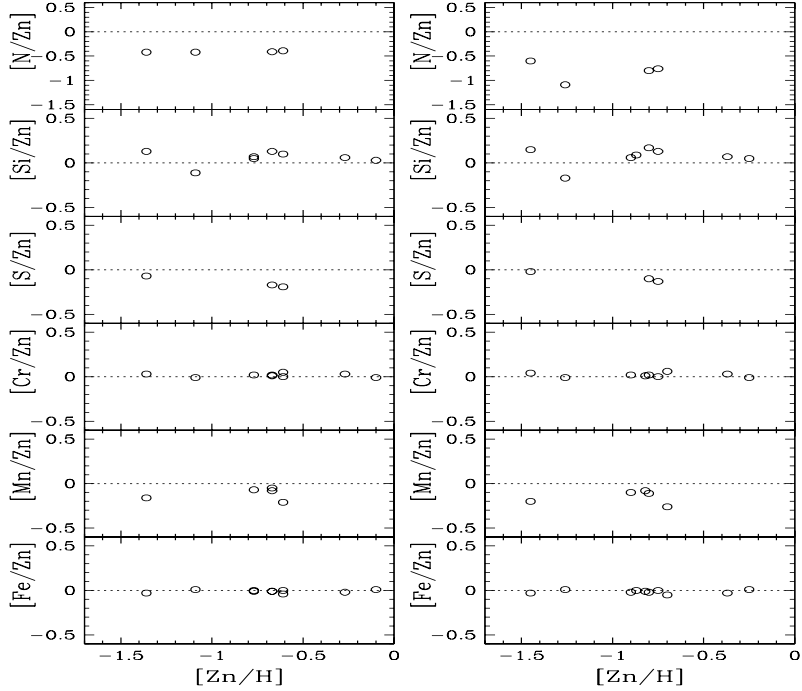,height=3.9in,width=4.5in}
}}
\caption{Inferred intrinsic abundance patterns of DLA systems for two different
choices of depletion patterns (see text).}
\end{figure}

Except for LSBCV, none of the previous discussions of DLA abundance
patterns included N in the analysis. LSBCV noted that the few then-existing
N abundance measurements cast doubt on the dust-depletion
interpretation of the Zn/Fe ratios in DLA systems. We have since obtained
additional N measurements to examine this issue more closely.
Figure 3 (left panels) shows the {\it inferred} 
intrinsic abundance patterns of DLAs after accounting for the elements 
in dust grains following the prescription by 
Vladilo (1997)\footnote{More N abundance measurements exist (Lu, Sargent,
\& Barlow 1998) but cannot be displayed in Figure 3 owing to lack of 
measurements of the corresponding Zn abundance.}.
The two DLA systems listed in Table 1 are not included in 
this analysis since neither shows any evidence for dust. 
The observed gas-phase abundance of N in the ISM is [N/H]$\simeq -0.2$ 
with little variations from sightline to sightline 
(Hibbert, Dufton, \& Keenan 1985). As was for other elements 
considered in the Vladilo model, we assumed in the above analysis
that the sub-solar N abundance in the ISM is due to dust depletion 
despite evidence against such an interpretation (e.g., Mathis 1996
and references therein; see next paragraph).
Interestingly, the results show neither
a solar pattern (N abundances being too low) nor a halo-star pattern (no
obvious $\alpha$ elements enhancement over Fe-group elements).
The deviations from the solar ratios in Si/Zn, S/Zn, and Mn/Zn, though small,
appear to be systematic.

Recent observations suggest that the {\it intrinsic} abundances of
C, N, O, and Kr in the present-day solar-neighborhood ISM are
about 2/3 solar rather than solar (Hibbert et al 1985;
Cardelli et al 1996; Sofia et al 1997;
Cardelli \& Meyer 1997), consistent with abundances
in young disk stars (see Snow \& Witt 1996 and references therein). 
This result has significant
implications for the compositions of Galactic  dust grains (Sofia, Cardelli,
\& Savage 1994; Snow \& Witt 1996; Mathis 1996),
which were previously derived
assuming solar abundances for the present-day ISM. To examine the
consequences of this result for the DLA abundances,
we re-derived the depletion pattern appropriate for warm disk 
clouds (Table 6 of Savage \& Sembach 1996) assuming an
intrinsic abundance of 2/3 solar for {\it all} elements,
and repeated the above analysis.
It turns out the results (Figure 3, right panels)
are little changed except for N. This is
because, while the depletion pattern for the warm diffuse clouds is
significantly changed with the adoption of the new reference abundances,
the scaled depletion pattern at the dust-to-metal ratios 
appropriate for the DLA systems is little affected. The result for N
changed significantly because N is no longer depleted with
the adoption of the new reference abundances. 

Can the inferred intrinsic abundance patterns shown in Figure 3
be understood with conventional chemical evolution scenarios?
Galactic halo stars show $\alpha$/Fe=2-3 times solar, 
which is thought to be the result
of SN II enrichment (Wheeler et al 1989).  It is generally believed that
significant contributions from SN Ia are required in order to 
reach solar $\alpha$/Fe ratio. However, 
the injection of large amounts of primary N into the interstellar
medium from intermediate mass stars (3-8 M$_{\odot}$) during the
AGB phase (Renzini \& Voli 1981), which must occur before making
SN Ia, should raise the N/Fe-group and N/$\alpha$ ratios to nearly solar values,
contrary to observations.
Having galactic winds carrying away some of the SN IIs ejecta
to reduce the $\alpha$-element abundances will not help since a
proportional amount of the Fe-group elements
will presumably be removed also, leaving the $\alpha$/Fe-group ratio in
the system unchanged.

\begin{table}[]
\begin{center}
\caption{Damped Ly$\alpha$ Systems Showing No Evidence For Dust}
\begin{tabular}{ccccccccc}
\hline
QSO &$z_{DLA}$   &[Fe/H] &[Si/Fe] &[Cr/Fe] &[Mn/Fe] &[Ni/Fe] &[Zn/Fe] \\
\hline
0454+0356$^a$ &0.8598 &$-1.00$ &...     &$-0.07$ &$-0.36$ &...     &$-0.08$ \\
1946+7658$^b$ &1.7382 &...$^b$ &$+0.25$ &$+0.13$ &$-0.41$ &$-0.44$ &$-0.09$ \\
\hline
\end{tabular}
\end{center}
$^a$ This system has log$N$(\textsc{H~i})=20.76. The abundances are 
from LSBCV and Steidel et al 1995.
$^b$ This system is inferred to be a damped Ly$\alpha$ system even though
the actual $N$(\textsc{H~i}) of the system is unknown (LSBCV). The abundances
are from our unpublished Keck observations except for Mn, which is from
Lu et al 1995. These values supersede those in LSBCV based on data with
lower S/N.
\end{table}

Given this situation,
it seems appropriate to examine other possibilities. We consider
the following alternatives: (1) the properties of dust grains in DLA
systems may be different from Galactic dust;
or (2) the nucleosynthetic history of DLA systems
may be different from that of the Milky Way galaxy.
The first possibility seems quite likely since dust grains in the
Magellanic Clouds are known from the extinction curves
to be different from Galactic dust and from each
other. More precise
determinations of the dust depletion properties in the Magellanic
Clouds' gas (e.g., Welty et al 1997) may provide useful clues.
There is also evidence that the second possibility may not be true
in the general sense. Two DLA systems are  known to have
near-solar Zn/Fe ratios (Table 1),  indicating the {\it absence} of dust 
by common criteria. Yet, the systems show
sub-solar Mn/Fe ratios and super-solar Si/Fe ratios; both 
characteristic of Galactic halo stars (Wheeler et al 1989). 
Interestingly, one of the systems has Ni/Fe ratio that
is a factor of 2.8 below solar, while halo stars
show nearly solar Ni/Fe ratios. This difference could be real, or,
more likely, it
may be due to a systematic error in the Ni~\textsc{ii} $f$-values
(LSBCV).  Detailed studies of such systems should provide important
clues to the intrinsic abundance patterns of DLA galaxies.

\acknowledgements{We thank
Art Wolfe and Jason X. Prochaska for providing
their abundance measurements in advance of publication.}

\end{document}